\documentclass[useAMS]{./mn2e}

\title[On the structure of the Sun and $\alpha$ Cen A and B]{On the structure of the Sun and $\alpha$ Centauri A and B in the light 
of seismic and non-seismic constraints}
 \author[M. Y{\i}ld{\i}z]{M. Y{\i}ld{\i}z$^{}$\thanks{E-mail:
mutlu.yildiz@ege.edu.tr},\\
Ege University, Department of Astronomy and Space Sciences, Bornova, 35100 \.Izmir, Turkey}

\input{./psfig}

\begin{document}

\date{Accepted 2005 December 15. Received 2005 December 14; in original form 2005 October 11}

\pagerange{\pageref{firstpage}--\pageref{lastpage}} \pubyear{2002}

\maketitle

\label{firstpage}

\begin{abstract}

{
The small separation ($\delta \nu_{01}$, $\delta \nu_{02}$ and $\delta \nu_{13}$) 
between the oscillations with low degree $l$ is 
dependent primarily on the sound speed profile within the stellar core, where nuclear evolution 
occurs. The detection of such oscillations for a star offers a very good opportunity
to determine the stage of its nuclear evolution, and hence its age. In this context,
we investigate the Sun and $\alpha$ Cen  A and B.
}
For 
$\alpha$ Cen  A and B, each of the small separations $\delta \nu_{01}$, 
$\delta \nu_{02}$ and $\delta \nu_{13}$ gives a different age. Therefore, in 
our fitting process, we also employ the second difference, defined as 
$\nu_{n,2}-2\nu_{n,1} + \nu_{n,0}$, which is 
$2\delta \nu_{01}-\delta \nu_{02}$.  In addition to this, we also use 
frequency ratio ($\nu_{n,0}/\nu_{n,2}$).  For the Sun, these expressions are
equivalent and give an age of about 4.9-5.0 Gyr. For $\alpha$ Cen  A and B, 
however, the small separation and the second difference give very different 
ages.  This conflict may be solved by the  detection of 
oscillation frequencies that can be measured much more precisely than the current frequencies.  
{
When we  fit the models to the observations, 
we find (i) $Z_0=0.020$, $t=3.50$ Gyr and $M_{\rm B}=1.006$M$_\odot$ from the small separations $\delta \nu_{01}$, 
$\delta \nu_{02}$ and $\delta \nu_{13}$ of $\alpha$ Cen B; and
(ii) a variety of solutions from the non-seismic constraints and $\delta \nu_{02}$ of $\alpha$ Cen A and B, in which
the masses of $\alpha$ Cen  A and B are slightly modified and the age of the system is about 5.2-5.3 Gyr. 
For $Z=0.025$, the closest masses we find to the observed masses are $M_{\rm B}$=0.922 $ \rm M_{\odot}$ and $M_{\rm A}$=1.115 $ \rm M_{\odot}$.
The differences between these masses and the corresponding observed masses are about 0.01 $ \rm M_{\odot}$.
}
\end{abstract}


\section{Introduction}
Microscopic or macroscopic, solid or fluid of any kind, every object oscillates at frequencies determined by
its structure. In many cases, the oscillations sound the structure of the visible and invisible parts 
of the object in which they are trapped. 
Among such objects are stars, particularly solar-like stars, which
oscillate in a variety of modes so that
one can infer physical conditions deep in the nuclear core (Tassoul 1980), which is not directly observable.  
In this context, Christensen-Dalsgaard (1988) developed a seismic Hertzsprung-Russel (HR) diagram in which
one can deduce the size and evolutionary phase  (age) of a star from its oscillation frequencies. 
In this diagram, the horizontal 
axis is so-called the large separation  between frequencies of consecutive oscillations with order $n$ 
($\Delta \nu_{nl}=\nu_{n,l}-\nu_{n-1,l}$) and 
the vertical axis is the small separation  
between frequencies of oscillations 
with different harmonics. As a result of nuclear evolution, the position of a star
in the seismic HR diagram changes with time. 
Roxburgh \& Vorontsov (2003, hereafter RV2003) argued that the 
ratio of the small separation to the large separation is more sensitive to time than the accurate value of 
the small separation, and proposed an alternative expression for it.  

In many respects, and particularly because of its high quality data, the Sun is an excellent object to study stellar interiors for
calibration of evolution code to be useds. 
In other words, our success in modelling interiors of stars
of different kinds depends on how successful we are in modelling the solar interior. The 
solar models with chemical composition given by Grevesse \& Sauval 
(1998) are  
in very good agreement with seismic inferences (Bahcall et al. 1995; Christensen-Dalsgaard et al. 1996; Gabriel \& Carlier 1997; Y{\i}ld{\i}z 2001). 
However, the agreement disappears if the recent solar chemical composition given by Asplund, Grevesse \& Sauval (2005, hereafter AGS2005)  
is used (see Basu et al. 2007 and references therein). Although uncertainties in the abundances of 
heavy elements are discussed by Pinsonneault \& Delahaye (2006) and found large enough to cover 
the old solar composition, an increase in diffusion coefficient (Guzik, Watson \& Cox 2005) and opacity (Bahcall et al. 2005) below the
convective zone are suggested to restore the agreement between the models and helioseismic constraints. 

In our previous paper on $\alpha$ Cen (Y{\i}ld{\i}z 2007), we obtained
8.9 Gyr for the age of the system from the  non-seismic constraints (NOS models in Y{\i}ld{\i}z 2007) and  about 5.6-5.9 Gyr 
from the seismic
constraints (SIS Models in Y{\i}ld{\i}z 2007; see also Miglio \& Montalban 2005).
There may be observational or theoretical reasons for this discrepancy. In this study, we investigate
how well core structures of the Sun and $\alpha$ Cen A and B are represented by different forms of small separations 
and how precise their fundamental properties are.
 
As main-sequence (MS) stars evolve their oscillation frequencies decrease because of 
the increase in their size. However, the amount of decrease in the frequencies depends also on how the sound speed 
changes in the cavity in which the oscillations are trapped.  As a result of nuclear evolution, the sound speed gradually decreases in the nuclear core,
and therefore frequencies of oscillations with $l=0$ are much more reduced than those are with $l=1$ and $l=2$.
Thus, the frequency ratio 
$\nu_{\rm n,0}$/$\nu_{\rm n,2}$ is a function of time and has in principle diagnostic potential as the small separation for information about the age of solar-like stars.
We also use this expression, despite its lower sensitivity to time than $\delta \nu_{02}$, in order to test if it gives the same age as the usual expressions (see Section 3).

In Y{\i}ld{\i}z (2007), the small separation $D_0$ is computed from $\delta \nu_{02}$ for $\alpha$ Cen A, and from $\delta \nu_{02}$ and $\delta \nu_{13}$ for $\alpha$ Cen B ($D_0=(\delta \nu_{02}/6+\delta \nu_{13}/10)/2$),
since $\delta \nu_{01}=(\nu_{n0}-(\nu_{n-1,1}+\nu_{n1})/2)$ values of models are quite different from the values found from the observed frequencies.
In this study, using in addition the seismic data of $\alpha$ Cen A given by Bedding et al. (2004, BK2004), we consider if it is possible to fit model values of $\delta \nu_{01}/2$, 
$\delta \nu_{02}/6$ and $\delta \nu_{13}/10$ one by one to the corresponding values inferred from the seismic data of the Sun and $\alpha$ Cen A and B.   

The remainder of this paper is organized as follows. In Section 2, the basic properties of solar models 
with old and recent chemical composition are presented.
The results are presented and discussed in
Section 3.
Finally, we make some 
concluding remarks in Section 4.

\section{Properties of the Solar Models}
   \begin{figure}
\centerline{\psfig{figure=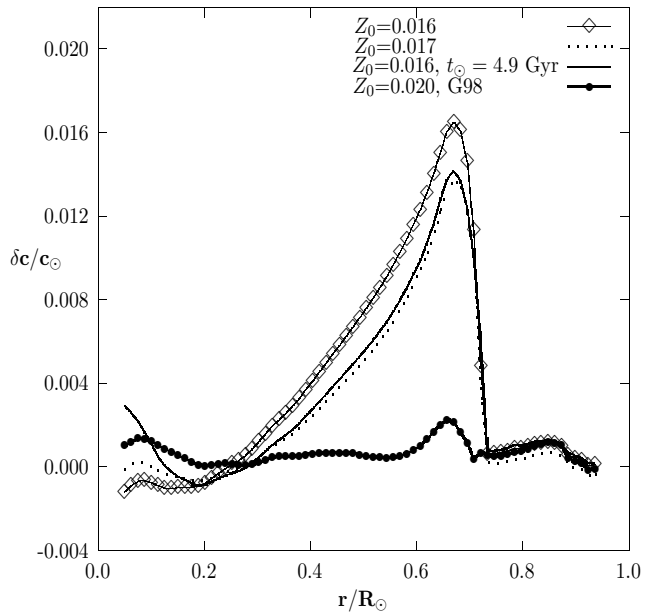,width=210bp,height=250bp}}
      \caption{ The relative sound speed difference between the solar models and the Sun is plotted with respect to the relative 
                radius. The first three models in the label are constructed with the chemical composition given in  AGS2005.
                The solid line with diamonds and dotted line show the solar models with $Z_0=0.016$ and $Z_0=0.017$, respectively. 
                The thick solid line represents the solar model with $Z_0=0.016$ and $ \rm t_\odot$ = 4.9 Gyr (see Section 3.4). The thick solid line with the filled circles 
                is for the solar model with the chemical composition given by Grevesse \& Sauval (1998), which is in good agreement with the helioseismic results.
}
              {\label{f1.4}}
   \end{figure}
\begin{table*}
\begin{center}
      \caption{ Properties of the solar models with the chemical composition given by AGS2005. 
The last line is for the observed values.
The uncertainty in $R_c$ is 0.001 $ \rm R_\odot$ (Basu \& Antia 1997).
The observed values of the surface Helium and metal abundances are taken from Basu \& Antia (1995) and AGS2005, respectively.
}
{\label{t1.1}}
{\begin{tabular}{ccccccccccr}
\hline
 $X_0$  & $Z_0$&$\alpha$&$\rho_c$& $T_c$ &$X_c$ &$ Z_s $  &$Y_s$&$R_c/ \rm R_\odot$&$  \rm t_\odot(Gyr)  ) $ \\  
\hline
0.70975 & 0.016& 1.820  & 146.78 & 15.65 & 0.355 & 0.0124 &0.244&0.733 & 4.6  \\  
0.70331 & 0.017& 1.832  & 147.19 & 15.73 & 0.348 & 0.0132 &0.249&0.730 & 4.6  \\  
0.71280 & 0.016& 1.850  & 151.16 & 15.72 & 0.340 & 0.0123 &0.241&0.731 & 4.9  \\  
0.70458 & 0.020& 1.889  & 152.54 & 15.63 & 0.335 & 0.0191 &0.246&0.710 & 4.6  \\  
 --- &---  & ---   & ---  & ---  & ---  & 0.0122 &0.246&0.713 & \\  

\hline
\end{tabular}}
\end{center}
\end{table*}

The characteristics of the code used in the construction of models for stellar interiors are given in Y{\i}ld{\i}z (2007).
Further details of the code are given in the references of that paper.
 
Models of $\alpha$ Cen  A and B required for our analysis are given in Y{\i}ld{\i}z (2007). We construct new  
solar models, because significant changes have occurred in our knowledge of the chemical composition of the Sun (AGS2005).   

The basic properties of the solar models with the recent solar composition given by AGS2005 are 
listed in Table 1. In the first row of this table the solar model with $Z_0=0.016$, which is in very good agreement with 
the observed chemical composition of the Sun, is presented: its surface helium and heavy element abundances are 0.244 and 0.0124,
respectively. However, the base radius of its convective zone ($0.733  \rm R_\odot$) is significantly greater than the value 
inferred from helioseismolgy ($0.713\pm0.001  \rm R_\odot$, Basu \& Antia 1997). In order to test the influence of initial value of heavy element abundance, another solar model is 
constructed with $Z_0=0.017$. This model
is given in the second row of Table 1. The base radius of its convective zone ($0.730  \rm R_\odot$) is slightly smaller than that of
the solar model with $Z_0=0.016$, but an agreement with the value inferred from helioseismolgy is not achieved. The age of these two models is
taken as 4.6 Gyr.  For later usage, we also construct a solar model with $Z_0=0.016$ and age = 4.9 Gyr (see Section 3.4).

In Fig. 1, the relative sound speed difference between these solar models and the Sun is plotted with respect to
the relative radius. The relative sound speed difference between the solar model with  $Z_0=0.016$ and the Sun is represented with diamonds.
The largest difference, about 1.7 \%, occurs in the region just below the base of the convective zone.
The larger the initial heavy element abundance, the smaller the relative sound speed difference
in the outer part of the radiative interior.
For $Z_0=0.017$, for example, the largest difference reduces to 1.4 \% (the dotted line). A very similar profile for the relative sound speed difference
occurs for the model with $Z_0=0.016$ and age = 4.9 Gyr (solid line). For comparison, the model with $Z_0=0.02$ and mixture given by 
Grevesse \& Sauval (1998) is also plotted in Fig. 1. It is in better agreement with the helioseismic results than the solar models with AGS2005.

\section{Structure of nuclear core and oscillation frequencies}
   \begin{figure}
\centerline{\psfig{figure=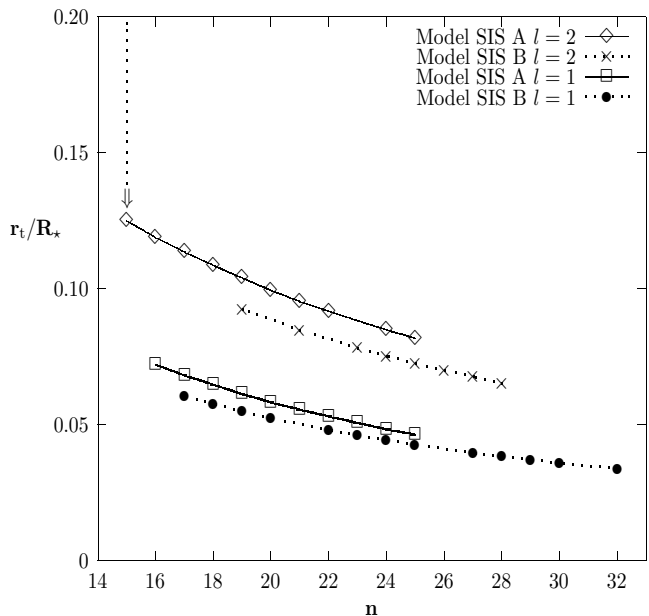,width=210bp,height=250bp}}
      \caption{The turning points of the observed modes in $\alpha$ Cen  A and B computed from SIS models of Y{\i}ld{\i}z (2007). 
}
              {\label{f1.3}}
   \end{figure}

The acoustic oscillations are trapped in the stellar cavity. Dimension of this cavity
depends on degree ($l$) and frequency of oscillations. Whereas the cavity for the oscillations with $l=0$
extends down to the center, the oscillations with higher $l$ have shallower cavities. The turning point (${r_{\rm t}}$)
of an oscillation with non-zero $l$ is approximately given as  
\begin{equation}
\frac{c_{\rm t}}{r_{\rm t}}=\frac{\omega}{\sqrt{l(l+1)}}
\end{equation}
where $\omega$ is the frequency of oscillation and $c(r_{\rm t})$ is the sound speed at the turning point. 
The turning points of the oscillations with observed frequencies of $\alpha$ Cen  A and B  are computed from SIS models
and plotted with respect to $n$ for the modes with $l=1$ and $l=2$ in Fig. 2. Whereas the  modes of $\alpha$ Cen  A (thin solid line 
with diamonds) and B (thin dotted line with $\times$s) with $l=2$ turns at nearly 
${r_{\rm t}/R_\star}\simeq 0.1$, the modes with $l=1$ (thick solid line with boxes 
for $\alpha$ Cen  A and dotted line with filled circles for $\alpha$ Cen  B) sink deeper into the 
nuclear core (${r_{\rm t}/R_\star}\simeq 0.05$).

   \begin{figure}
\centerline{\psfig{figure=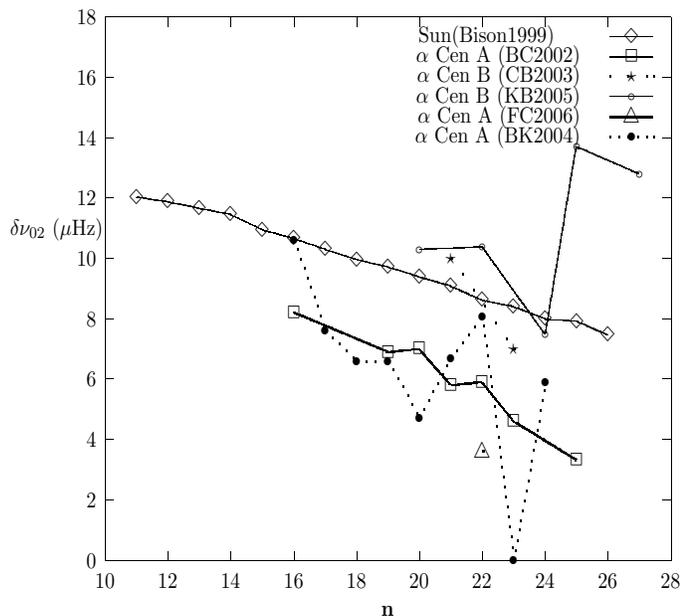,width=210bp,height=250bp}}
      \caption{ The small separation $\delta \nu_{02}$ between the acoustic oscillations of the Sun (diamonds), the components of $\alpha$ Cen A (boxes for BC2002, filled circles for BK2004 and triangle for FC2006) and B (circles for 
Kjeldsen et al. and star for Carrier \& Bourban 2003) as a function of $n$.  
}
              {\label{f1.4}}
   \end{figure}
   \begin{figure}
\centerline{\psfig{figure=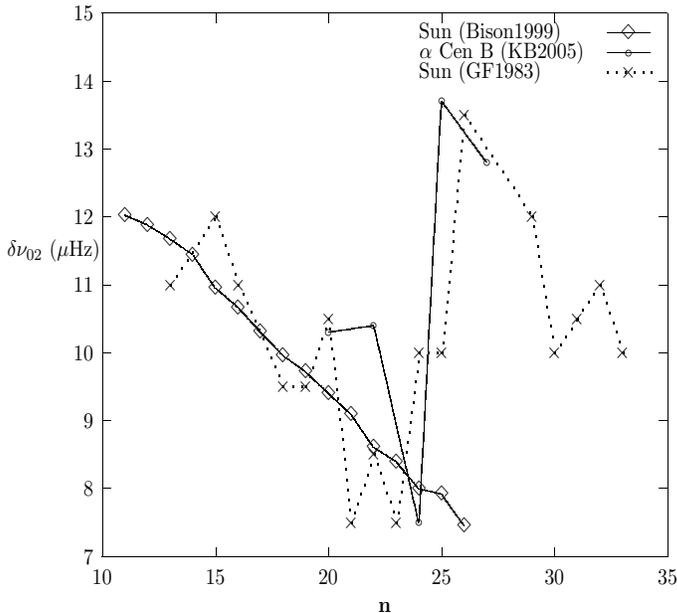,width=210bp,height=250bp}}
      \caption{ Comparison of the small separation $\delta \nu_{02}$ computed from the old (dotted line with $\times$s) and new (diamonds) acoustic oscillations of the Sun. $\delta \nu_{02}$ of $\alpha$ Cen  B (circles; KB2005) is also plotted. 
}
              {\label{f1.4}}
   \end{figure}

Whereas the small separation is a sensitive function of physical conditions in central regions, where the nuclear evolution 
occurs, the large 
separation, $ \Delta \nu_{0}=\nu_{n,0}-\nu_{n-1,0}$,  is a measure of the mean density. 
The small separation between frequencies (Christensen-Dalsgaard 1988),
$\delta \nu_{02}=\nu_{n,0}-\nu_{n-1,2}$,
is plotted as a function of $n$ in Fig. 3, using the observed frequencies of the Sun (thin solid line with diamonds; Chaplin et al. 1999),
 $\alpha$ Cen  A (thick solid line with boxes; Bouchy \& Carrier 2002, BC2002) and $\alpha$ Cen B (thin solid line with circles; Kjeldsen et al. 2005, 
hereafter KB2005). 
Its $n$ dependence is very similar to a straight line for the Sun and  $\alpha$ Cen  A (BC2002). Although its average value is about 
$5.6 ~\mu$Hz for $\alpha$ Cen  A, the difference between its maximum and minimum values is about $5 ~\mu$Hz. In other words, the small separation has 
a strong dependence on $n$. Also shown in Fig. 3 are the small separations of $\alpha$ Cen A (triangle) and B (stars) computed from the observations of 
Fletcher et al. (2006, FC2006) and Carrier \& Bourban (2003, CB2003), respectively. 
We note 
that the $\delta \nu_{02}$ from FC2006 is significantly smaller than that of BC2002.  
The dotted line with filled circles in Fig. 3 is for $\delta \nu_{02}$ from BK2004.
Although the difference between its maximum and minimum values is about $10 ~\mu$Hz, its mean value ($6.3 ~\mu$Hz) is close to the value found from BC2002 ($5.6 ~\mu$Hz).

{ 
In Fig. 3, $\delta \nu_{02}$ of the Sun (Chaplin et al. 1999) is much smoother than the corresponding values for $\alpha$ Cen  A and B. 
However, for the less accurate solar data of the early stages of helioseismology this was not the case.
In Fig. 4,  $\delta \nu_{02}$ of the old (Grec et al. 1983; GF1983; dotted line with $\times$s) and the recent (Chaplin et al. 1999) data of the Sun are plotted with respect to $n$.
For small values of $n$, despite the scattering of the Grec et al. (1983) data,  the two data sets are in agreement. However, for $n\geq 25$, there is a significant difference. 
Grec et al. (1983) state that the frequency resolution is 2 $\mu$Hz. The error in frequencies found by Chaplin et al. (1999) is about 0.05 $\mu$Hz for $n\leq 22$ and increases very rapidly
for larger values of $n$: for $n=26$, for example, it is about 0.3 $\mu$Hz.
We also note in Fig. 4 the similarity between the old data of the Sun and the data of $\alpha$ Cen B.    
As our skill in detecting the seismic properties of $\alpha$ Cen A and B develops, as in the solar case, we may obtain much more smooth variation of $\delta \nu_{02}$ with respect to $n$.

On the theory side, however, uncertainty arises from the fact that model frequencies are computed assuming an adiabatic process for the oscillations (Christensen-Dalsgaard \& Thompson 1997). 
The adiabatic approximation is valid for almost the entire interior, except for the near-surface regions.  
Therefore, the perturbation should also be applied to the energy equation, but the perturbation of energy equation is highly uncertain, 
at least for the convective flux (Christensen-Dalsgaard et al. 1996).
However, use of either the difference between or the ratio of frequencies for comparison is a good way to minimize the effects of such a troublesome problem.

}

\subsection[]{Sound speed profiles of the Sun and $\alpha$ Cen  A and B and the time variation of frequencies}
   \begin{figure}
\centerline{\psfig{figure=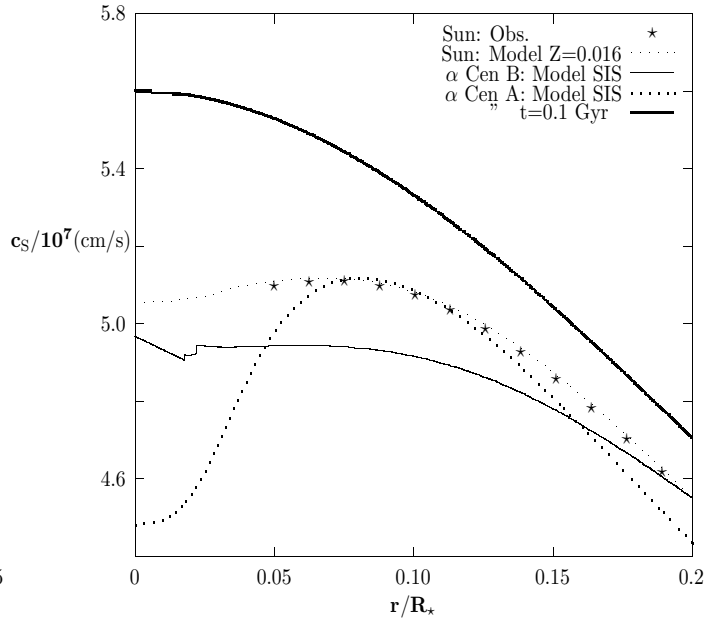,width=210bp,height=250bp}}
      \caption{Sound speed throughout the Sun and $\alpha$ Cen  A and B with respect to the relative radius near the central regions. 
}
              {\label{f1.3}}
   \end{figure}
The variation of stellar oscillation frequencies with time is primarily a result of  global expansion of the star like a heated ball. 
The secondary effect on the time dependence of frequencies arises from the 
fact that sound speed profile changes throughout the star. For the secondary effect, the most significant change occurs within the nuclear core.
Therefore, this secondary effect is important for modes with low degree. 
Whereas variation of the large separation with time is a good measure of the first effect, the variation of the small separation with time
is the result of the secondary effect. In Fig. 5, 
the sound speed profiles in the central regions of the Sun (thin dotted line, Basu et al. 1997), the solar model with $Z_0=0.016$ (stars),
SIS models of $\alpha$ Cen  A (thick dotted line) and B (thin solid line) are plotted with respect to the relative radius. 
For comparison, the sound speed (thick solid line) of $\alpha$ Cen  A (SIS model) near the zero-age main-sequence (ZAMS) 
($t=0.1$ Gyr) is also plotted. As stated above, the most significant difference between the sound speed profiles occurs in 
the nuclear core: as hydrogen is converted to helium, the mean molecular
weight increases and consequently the sound speed at the center drops from $5.6\times 10^7$ to $4.5\times 10^7$ cm s$^{-1}$.

The oscillations with $l=1$
are influenced much more than the oscillations with $l=2$ because of this variation in sound speed profile with time.
It is well known that the turning point of oscillations depends mainly on the degree of oscillations. Whereas the turning points of 
oscillations with $l=1$ are about $r_{\rm t}=0.05 R_\star$, the mean value for the turning points of
oscillations with $l=2$ is about $r_{\rm t}=0.10 R_\star$. As the nuclear evolution proceeds the sound speed decreases,
and consequently the oscillations with $l= 1$ and $l= 2$ sink deeper, according to equation (1). From the ZAMS to the present time, the decrease in sound 
speed at point $r_{\rm t}=0.05 R_\star$ with time 
for $\alpha$ Cen A, for example, is about $0.6~\times 10^7$ cm s$^{-1}$ (Fig. 5), 
at $r_{\rm t}=0.10 R_\star$, however, the change with time is three times smaller ($0.2~\times 10^7$ cm s$^{-1}$) than this value. 
This means that the increase in the path of the oscillations with $l=1$ is much more than that of the oscillations with $l=2$. 
Therefore, the decrease in the frequencies of the oscillations with $l=1$ as a result of nuclear evolution is much greater than that with $l=2$.

\subsection[]{The second difference defined as $\nu_{n,0}-2\nu_{n,1}+\nu_{n,2}$ }
   \begin{figure}
\centerline{\psfig{figure=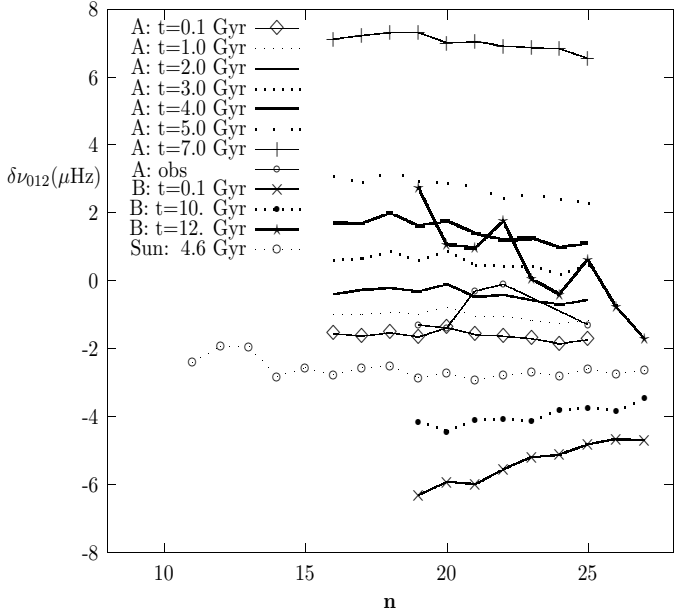,width=210bp,height=250bp}}
      \caption{ $\delta \nu_{012}$ as a function of n for $\alpha$ Cen A at various ages from the ZAMS to about the TAMS. 
                For comparison, $\delta \nu_{012}$ of the solar model ($Z_0=0.016$ and age=4.6 Gy) and the SIS models of 
                $\alpha$ Cen B are also plotted.
}
              {\label{f1.4}}
   \end{figure}
   \begin{figure}
\centerline{\psfig{figure=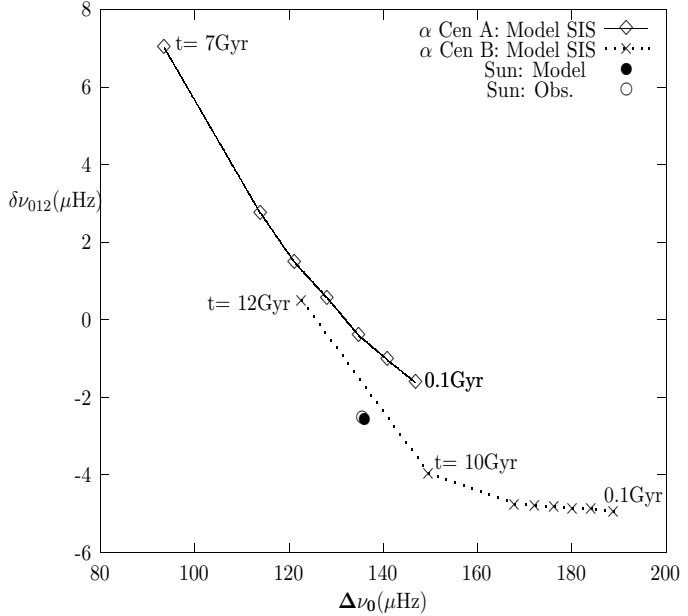,width=210bp,height=250bp}}
      \caption{ The evolutionary tracks of $\alpha$ Cen A and B in the alternative seismic HR Diagram. For comparison, the observed position of the Sun and 
the position found from the solar model with $Z_0=0.016$ are also plotted.
}
              {\label{f1.4}}
   \end{figure}

As discussed above, oscillation frequencies of a star are a decreasing function of time, 
because of the increase of its size and sound speed profile over time as a result of nuclear evolution. 
Christensen-Dalsgaard (1988) and RV2003
use the frequencies of oscillations with different $n$ and $l$ for the expression for 
the small separation.
We use also the second difference derived from the frequencies of mode with the same $n$:
\begin{equation}
\delta \nu_{012} = (\nu_{\rm n,0}- 2 \nu_{\rm n,1}+\nu_{\rm n,2})
\end{equation}
The second difference is the difference between $2\delta \nu_{01}$ and $\delta \nu_{02}$.
According to the asymptotic relation, $\delta \nu_{012}$ is very small, just like $\delta \nu_{02}$.
In Fig. 6,  $\delta \nu_{012}$ is plotted with respect to $n$ at various evolutionary 
times of $\alpha$ Cen  A, from the ZAMS to the terminal-age-MS (TAMS).  Close to the ZAMS (thin solid line with diamonds),
$\delta \nu_{012}$ is negative and varies between $-2.0~\mu$Hz and $-1.5~\mu$Hz.
As the model evolves, $\delta \nu_{012}$ becomes positive at an age of nearly half of the 
MS life time of the star, and reaches an average value of about $7.0~\mu$Hz at age=7 Gyr. 
$\delta \nu_{012}$ computed from the observed frequencies is also plotted in Fig. 6. 
From the comparison of the model and the observational $\delta \nu_{012}$,      
contrary to the result of $\delta \nu_{02}=\nu_{n,0}-\nu_{n,2}$, we deduce that $\alpha$ Cen  A must be  
in the early phase of its MS evolution. 

The solar model with $Z_0=0.016$ and age=4.6 Gyr is also plotted in Fig. 6 (dotted line with circles).
As for $\alpha$ Cen  A, $\delta \nu_{012}$ of the Sun is also independent of $n$. However, the model value of $\delta \nu_{012}$ for $\alpha$ Cen B
is dependent on $n$ near the ZAMS (solid line with $\times$s) and the TAMS (thick solid line with stars) and has less $n$ dependence between these two cases (dotted line with filled circles for a model with age=10 Gyr) . 

For the Sun, for which we have high quality of seismic data, average value of $\delta \nu_{012}$ derived from seismic data
over $n$ is $-2.50~\mu$Hz and it is $-2.57~\mu$Hz for the solar model with $Z_0=0.016$ and age=4.6 Gyr (dotted line with circles in Fig 6);
that is, the observed value is 
slightly greater than the model value (the same result is obtained from 
$\delta \nu_{02}=\nu_{n,0}-\nu_{n,2}$; see Section 3.4).   

In Fig. 7, average value of $\delta \nu_{012}$ over n is plotted against $\Delta \nu_{0}$ during the 
MS evolution (SIS model) of $\alpha$ Cen  A (solid line with diamonds) and B (dotted line with $\times$s). The ZAMS is represented 
by the symbol closest to the legend '0.1 Gyr'. The next closest symbol corresponds to $t=1$ Gyr. After that symbol, time increment 
between two symbols without legend is 1 Gyr. In the early phase of MS evolution, $\delta \nu_{012}$ of 
$\alpha$ Cen  B remains almost constant, whereas $\delta \nu_{012}$ of $\alpha$ Cen  A increases very rapidly. In contrast to 
$\delta \nu_{02}$, again, $\delta \nu_{012}$ of the massive star ($\alpha$ Cen  A) is much more sensitive function of time
than that of the low-mass star ($\alpha$ Cen  B). 

\subsection[]{Ratio of frequencies}
   \begin{figure}
\centerline{\psfig{figure=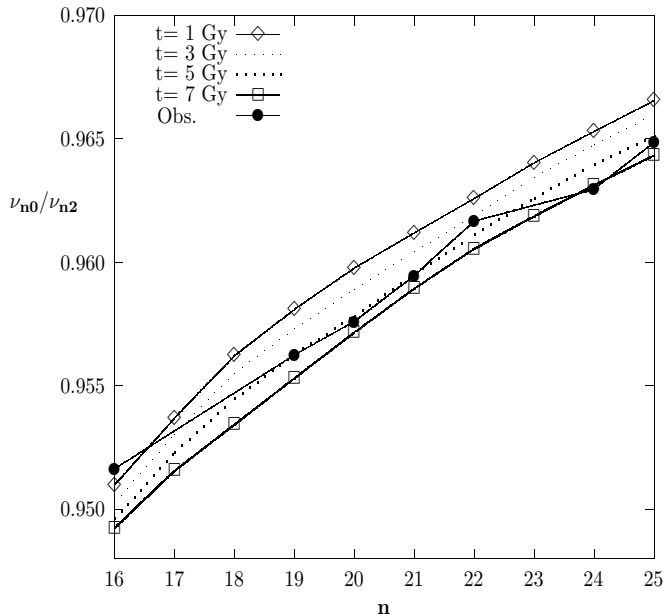,width=210bp,height=250bp}}
      \caption{ The ratio $ \nu_{n,0}/\nu_{n,2}$ from SIS model of $\alpha$ Cen  A with respect to n. For comparison, the 
observed ratio is also plotted (filled circles).
}
              {\label{f1.467}}
   \end{figure}

Although the ratio $\nu_{\rm n,0}$/$\nu_{\rm n,2}$ is not a more sensitive function of time than $\delta \nu_{\rm 02}$, 
we want to test whether it gives the same age as $\delta \nu_{\rm 02}$.
%
%
In Fig. 8, ratio of $\nu_{\rm n,0}$ to $\nu_{\rm n,2}$ for $\alpha$ Cen  A is plotted as a function of $n$ at 1 Gyr 
(thin solid line with diamonds), 3 Gyr (thin dotted line), 5 Gyr (thick dotted line) and 7 Gyr (thick solid line with boxes).
The decrease in the ratio with time is noticeable. For comparison, the observed ratio is also plotted 
(solid line with filled circles). The observed ratio is in agreement with the curve of 5 Gyr, 
except for the ratio for $n=16$.     
 

\subsection{Age and metallicity of the Sun from the seismic constraints}
The observed value of $\delta \nu_{02}$ for the Sun is found to be $9.84 \pm 0.1~\mu$Hz from the seismic 
data of the Bison group. The value of $\delta \nu_{02}$ for a solar model with $Z_0=0.016$ is, however, 10.11 $~\mu$Hz; 
that is, it is slightly greater than the observed value. 
For the solar model with $Z_0=0.017$, $\delta \nu_{02}=10.08 ~\mu$Hz.  The value of $\delta \nu_{02}$ for a solar model with $Z_0=0.016$
and age=4.9 Gyr, $9.86~\mu$Hz,  is in very good agreement with the observed value. 
From the ratio of frequencies ($\nu_{\rm n,0}/\nu_{\rm n,2}$),
we confirm that the same age (4.9 Gyr) is more suitable than 4.6 Gyr. 
Similarly, from the comparison of the observed value of $\delta \nu_{012}$ ($2.50~\mu$Hz) with that of the solar model 
with $Z_0=0.016$ and age=4.6 Gyr ($2.57~\mu$Hz), we find that the solar model with 4.9 Gyr is in better agreement with 
the helioseismic data. 

%
%
%
%
%
%
%

\subsection{Are values of $\delta \nu_{01}/2$, $\delta \nu_{02}/6$ and $\delta \nu_{13}/10$ compatible with each other}

\begin{table*}
\begin{center}
      \caption{ $D_{01}$, $D_{02}$ and $D_{13}$  of the solar models and models of $\alpha$ Cen  A and B.
}
{\label{t1.3}}
{\begin{tabular}{l|cccrrrl}
\hline
Star&  $t(Gyr)$ & $Z_0$   & $X_0$  &     $D_{01}(\mu$Hz)  &     $D_{02}(\mu$Hz)& $D_{13}(\mu$Hz) & Model   \\
\hline
Sun &   4.60 & 0.0160 & 0.70975 & 1.879  &  1.686 &  1.739 &    \\
    &   4.60 & 0.0170 & 0.70331 & 1.870  &  1.679 &  1.734& \\
    &   4.60 & 0.0190 & 0.69080 & 1.858  &  1.659 &  1.713& \\
    &   4.90 & 0.0160 & 0.71280 & 1.825  &  1.644 &  1.701& \\
    &   4.90 & 0.0200 & 0.68806 & 1.825  &  1.613 &  1.676&  \\
    &   4.93 & 0.0170 & 0.70665 & 1.837  &  1.632 &  1.695& \\
    &   4.98 & 0.0165 & 0.71039 & 1.835  &  1.628 &  1.691& \\
Obs C99 &   --   &   --   &   --   & 1.813&  1.640 & 1.679 &  \\ 
Obs B07 &   --   &   --   &   --   &  --  &  1.653 & 1.683 &  \\ 
     &        &        &           &  --  &$\pm$ 0.005&$\pm$ 0.007  & \\ 
\hline
$\alpha$ Cen A 
    &   5.70 & 0.0322 & 0.669 & 2.41  &  0.93 &  0.76& SIS\\
    &   8.88 & 0.0328 & 0.703 & 2.53  &  0.49 &  0.37& NOSIS\\
    &   5.90 & 0.0230 & 0.713 & 2.24  &  0.94 &  0.82& SIS23\\
    &   3.50 & 0.0200 & 0.756 & 3.01  &  1.44 &  1.11& SIS2\\
    &   3.50 & 0.0200 & 0.732 & 2.73  &  1.41 &  1.14& SIS2p\\
Obs BK2004 
    &   --   &   --   &   --   & 1.41  &  1.05 & 1.10 &  \\ 
Obs BC2002 &   --   &   --   &   --   & 0.98  &  0.99 &  --   &  \\ 
Uncertainty    &   --   &   --   &   --   & 0.12  &  -- &  --   &  \\ 
\hline
$\alpha$ Cen B
    &   5.70 & 0.0322 & 0.669 & 0.95  &  1.53 &  1.86& SIS\\
    &   8.88 & 0.0328 & 0.703 & -0.05  &  0.92 &  1.44& NOSIS\\
    &   5.90 & 0.0230 & 0.713 & 0.98  &  1.49 &  1.71& SIS23\\
    &   3.50 & 0.0200 & 0.756 & 1.83  &  1.82 &  1.81& SIS2\\
    &   3.50 & 0.0200 & 0.732 & 1.60  &  1.82 &  1.88& SIS2p\\
Obs KB2005 
    &   --   &   --   &   --   & 2.26  &  1.69 & 1.67 &  \\ 
Uncertainty    &   --   &   --   &   --   & 0.26  &  0.06 & 0.07 &  \\ 
\hline
\end{tabular}}
\end{center}
\end{table*}

The small separation $D_0$ can be derived from $\delta \nu_{01}$, $\delta \nu_{02}$ and $\delta \nu_{13}$, if we have enough observational data:

\begin{equation}
D_0=(D_{01}+D_{02}+D_{13})/3=(\delta \nu_{01}/2+\delta \nu_{02}/6+\delta \nu_{13}/10)/3.
\end{equation}
As emphasized above,
in Y{\i}ld{\i}z (2007), $D_{\rm 0}$ is computed from $\delta \nu_{02}$ for $\alpha$ Cen A and from $\delta \nu_{02}$ and $\delta \nu_{13}$ for 
$\alpha$ Cen B. Its values inferred from the observed seismic frequencies ($D_{\rm 02,obs}$ and $D_{\rm 13,obs}$) and models with 
an age of about 5.7-5.9 Gyr are in good agreement. The values of $D_{\rm 01,obs}=\delta \nu_{01}/2$  
of both $\alpha$ Cen A and B are not used in the analysis in Y{\i}ld{\i}z (2007), because they are not in agreement with the $D_{\rm 01}$ values derived from SIS models 
and are not compatible with $D_{\rm 02,obs}$ and $D_{\rm 13,obs}$ (see Table 2). In fact, the disagreement between the values of $\delta \nu_{012}$ computed from the models 
and the values inferred from the seismic data (see section 3.2) arises from this inharmoniousness. 
Such an inharmoniousness does not exist for the Sun, and  we may hope to remove it with observational seismic data for $\alpha$ Cen A and B that are much more precise than the present data. 

We first consider the solar data and try to fit $D_{01}$, $D_{02}$ and $D_{13}$ of the solar models to $D_{\rm 01,obs\odot}$, $D_{\rm 02,obs\odot}$ and $D_{\rm 13,obs\odot}$.
In Table 2, $D_{01}$, $D_{02}$ and $D_{13}$ of the solar models and models of $\alpha$ Cen A and B are listed. In the first three rows the solar models with an age of 
4.60 Gyr and with different $Z_0$ values are given. $D_{01}$, $D_{02}$ and 
$D_{13}$ of these models are greater than $D_{\rm 01,obs\odot}$, $D_{\rm 02,obs\odot}$ and $D_{\rm 13,obs\odot}$ (given in the eighth row), respectively.
For comparison, the values of 
$D_{\rm 02,obs\odot}$ and $D_{\rm 13,obs\odot}$ derived from the recent seismic solar data of Basu et al. (2007) and their uncertainties are given in ninth and tenth rows, respectively.  
In the fourth row, the solar model with an age of 4.9 Gyr and $Z_0=0.016$ is given. Its $D_{02}$ is the same as $D_{\rm 02,obs}$, and $D_{01}$ and $D_{13}$ are very close to
$D_{\rm 01,obs\odot}$ and $D_{\rm 13,obs\odot}$, respectively. In order to find the optimum values of $Z_0$ and an age for the Sun we write three equations for each $D_{ll'}$,  
for instance,
\begin{equation}
D_{\rm 01,obs\odot}=D_{\rm 01,ref}+\frac{\Delta D_{\rm 01}}{\Delta Z_0}\delta Z_0 +\frac{\Delta D_{\rm 01}}{\Delta t}\delta t,
\end{equation}
and solve them for $Z_0$ and age ($t$). We compute the required derivatives such as given in equation (4) from the solar models discussed above and find the solution:
$Z_0=0.017$ and $t=4.93$ Gyr. The solar model with these values is given in the sixth row of Table 2. We use this model as the reference model and obtain another solution: 
$Z_0=0.0165$ and $t=4.98$ Gyr. The solar model with the new results is presented in the seventh row. 
$D_{ll'}$ of these solar models with about $Z_0=0.0165-0.0170$ and age about $4.9-5.0$ Gyr 
are very similar values and are compatible with the results inferred from helioseismolgy.

In this context, we also consider $\alpha$ Cen A and B. Their models from Y{\i}ld{\i}z (2007) are given in the first three rows of their corresponding part in Table 2.  
For our analysis of $\alpha$ Cen A and B, we write again three equations similar to equation (4) for $D_{ll'}$ for each star but with three unknowns. 
In addition to $Z_0$ and $t$ , we assume that the mass of the stars are also unknown.
We obtain separate solutions for $\alpha$ Cen A and B. For $\alpha$ Cen B, $Z_0=0.020$, $t=3.50$ Gyr and $M_{\rm B}=1.006$M$_\odot$.
For $\alpha$ Cen A, we obtain $t=3.50$ Gyr and $M_{\rm A}=1.19$M$_\odot$. However, value of $Z_0$ for the solution for $\alpha$ Cen A is not a reasonable  value and therefore we use the same value of $Z_0$
as obtained for $\alpha$ Cen B. Models of $\alpha$ Cen A and B (SIS2) with these values are given in the fourth row of their corresponding parts in Table 2. Although these values are the optimum values
in order to fit model values of $D_{ll'}$ to $D_{ll',\rm obs}$, there are still significant differences between $D_{01}$ and $D_{01,obs}$ of each star. We construct similar models for 
$\alpha$ Cen A and B with the same values but the masses ($M_{\rm A}=1.15$M$_\odot$ and $M_{\rm B}=0.97$M$_\odot$) are average of the masses we find and the masses given by Pourbaix et al. (2002).
These models (SIS2p) are presented in the fifth row of the corresponding parts of $\alpha$ Cen A and B in Table 2. $D_{01}$ values of the SIS2 and SIS2p models are significantly different. 
However, there is no simultaneous agreement between the model and the observed values of $D_{ll'}$ for $\alpha$ Cen A and B as in the case of the Sun.

\section{Masses and age of $\alpha$ Cen A and B}
\begin{figure}
\centerline{\psfig{figure=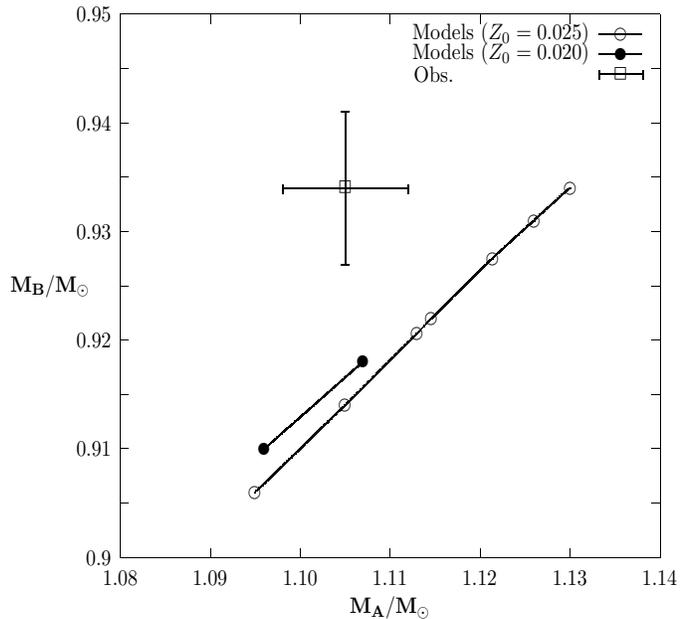,width=210bp,height=250bp}}
      \caption{ By modification of masses of $\alpha$ Cen A and B, full agreement between the observational 
                and theoretical results is achieved. However, there is no unique solution. Various
                combinations of $M_{\rm A}$ and $M_{\rm B}$ are possible. 
                The mass of $M_{\rm B}$ required for the full agreement (circles) is plotted with respect 
                to $M_{\rm A}$ for $Z_0=0.025$.  The solutions for $Z_0=0.020$ are represented by filled circles.
                The observational value is also shown (boxes).
}
              {\label{f1.4}}
   \end{figure}
\begin{table*}
\begin{center}
      \caption{ Models with slightly modified masses, that are in full agreement with observations.
}
{\label{t1.1}}
{\begin{tabular}{lccccccccccr}
\hline
Star& $M/ \rm M_{\odot}$&t(Gyr) & $Z_0$ & $\alpha$ & $X_0$ & $L/ \rm L_{\odot}$ & $R/ \rm R_{\odot}$ & $T_{\rm eff}$(K) & $D_{02}(\mu$Hz)& $\Delta \nu_{0}(\mu$Hz)  \\
\hline
A     & 1.097 &  5.15  & 0.025 &  1.588    & 0.68950  & 1.544  & 1.224 & 5824 & 1.072   & 106.0\\   
B     & 0.907 &  5.15  & 0.025 &  1.550    & 0.68950  & 0.507 & 0.864  & 5243 & 1.679   & 162.3\\   
\hline

A     & 1.113 &  5.30  & 0.025 &  1.608    & 0.70133  & 1.542  & 1.228 & 5809 & 1.049   & 106.1\\   
B     & 0.921 &  5.30  & 0.025 &  1.644    & 0.70133  & 0.508 & 0.863  & 5252 & 1.670   & 163.8\\   
\hline
A     & 1.115 &  5.24  & 0.025 &  1.611    & 0.70256  & 1.536  & 1.225 & 5812  & 1.039   & 106.6\\   
B     & 0.922 &  5.24  & 0.025 &  1.648    & 0.70256  & 0.505  & 0.862  & 5248 & 1.679   & 164.22\\   
\hline

A     & 1.130 &  5.25  & 0.025 &  1.634    & 0.71180  & 1.539  & 1.224 & 5816  & 1.067   & 107.4\\   
B     & 0.934 &  5.25  & 0.025 &  1.694    & 0.71180  & 0.507 & 0.863  & 5248  & 1.687   & 164.7\\   
\hline
      
A     & 1.096 &  5.35  & 0.020 &  1.575    & 0.71830  & 1.538  & 1.223 & 5855  & 1.086   & 106.0\\   
B     & 0.910 &  5.35  & 0.020 &  1.635    & 0.71830  & 0.508  & 0.863  & 5250  & 1.681   & 162.8\\  
\hline
      
A     & 1.107 &  5.37  & 0.020 &  1.582    & 0.72504  & 1.552  & 1.228 & 5819  & 1.075   & 105.7\\   
B     & 0.918 &  5.37  & 0.020 &  1.637    & 0.72504  & 0.507  & 0.866 & 5239  & 1.680   & 162.6\\   
\hline
A (obs.)     & 1.105 &  --    & --   &    --    & --    & 1.544  & 1.224 & 5824  & 1.05   & 105.6\\   %
B (obs.)     & 0.934 &  --    & --   &  --      & --    & 0.507  & 0.863  & 5250   & 1.69   & 161.5\\   %
\hline

\end{tabular}}
\end{center}
\end{table*}

\begin{table*}
\label{ta2}
\caption{
Adiabatic oscillation frequencies of the best fitted models for $\alpha$ Cen A and B with masses $1.115$ $\rm M_{\odot}$ and $0.922$ $\rm M_{\odot}$, respectively.  
}
$
\begin{array}{ccccc|cccc}
\hline 
            \noalign{\smallskip}
    &         &  \alpha ~{\rm~Cen~A}&   &         &         & \alpha  ~{\rm~Cen~B}&     &           \\
n   & \nu_{n0}& \nu_{n1}& \nu_{n2}& \nu_{n3}& \nu_{n0}& \nu_{n1}& \nu_{n2}& \nu_{n3}  \\
                    \noalign{\smallskip}
                    \hline
                    \noalign{\smallskip}

   10 & 1219.56 & 1267.70 & 1320.00 & 1108.49 & 1844.75 & 1924.54 & 1995.22 & 1676.04\\
   11 & 1327.68 & 1375.06 & 1426.31 & 1219.56 & 2010.85 & 2089.51 & 2160.28 & 1844.75\\
   12 & 1433.86 & 1480.79 & 1532.34 & 1327.68 & 2175.42 & 2253.87 & 2324.48 & 2010.85\\
   13 & 1539.96 & 1587.38 & 1639.11 & 1433.86 & 2339.09 & 2418.09 & 2489.06 & 2175.42\\
   14 & 1646.69 & 1694.54 & 1746.41 & 1539.96 & 2503.03 & 2581.58 & 2652.87 & 2339.09\\
   15 & 1753.84 & 1801.49 & 1852.86 & 1646.69 & 2666.40 & 2744.84 & 2815.82 & 2503.03\\
   16 & 1860.08 & 1907.62 & 1958.69 & 1753.84 & 2828.74 & 2907.42 & 2978.97 & 2666.40\\
   17 & 1965.66 & 2013.16 & 2064.48 & 1860.08 & 2991.41 & 3070.09 & 3142.26 & 2828.74\\
   18 & 2071.30 & 2119.40 & 2171.02 & 1965.66 & 3154.33 & 3233.53 & 3305.99 & 2991.41\\
   19 & 2177.64 & 2226.17 & 2278.23 & 2071.30 & 3317.60 & 3397.04 & 3470.18 & 3154.33\\
   20 & 2284.65 & 2333.49 & 2385.51 & 2177.64 & 3481.40 & 3560.66 & 3634.05 & 3317.60\\
   21 & 2391.67 & 2440.74 & 2492.85 & 2284.65 & 3644.93 & 3724.53 & 3798.19 & 3481.40\\
   22 & 2498.74 & 2547.79 & 2599.94 & 2391.67 & 3808.69 & 3888.42 & 3962.70 & 3644.93\\
   23 & 2605.60 & 2654.97 & 2707.21 & 2498.74 & 3972.89 & 4052.78 & 4127.39 & 3808.69\\
   24 & 2712.60 & 2762.14 & 2814.69 & 2605.60 & 4137.29 & 4217.50 & 4292.41 & 3972.89\\
   25 & 2819.87 & 2869.70 & 2922.25 & 2712.60 & 4301.98 & 4382.13 & 4457.40 & 4137.29\\
   26 & 2927.18 & 2977.24 & 3029.99 & 2819.87 & 4466.71 & 4546.95 & 4622.34 & 4301.98\\
   27 & 3034.68 & 3084.85 & 3137.60 & 2927.18 & 4631.36 & 4711.65 & 4787.36 & 4466.71\\
   28 & 3142.05 & 3192.45 & 3245.21 & 3034.68 & 4796.13 & 4876.35 & 4952.35 & 4631.36\\
   29 & 3249.41 & 3299.82 & 3352.64 & 3142.05 & 4960.90 & 5041.26 & 5117.37 & 4796.13\\
   30 & 3356.62 & 3407.17 & 3459.85 & 3249.41 & 5125.69 & 5206.10 & 5282.43 & 4960.90\\

                    \noalign{\smallskip}
                    \hline
        \end{array}
        $
                                                                                                                                                                                  
        \end{table*}

An agreement between the ages of $\alpha$ Cen A and B derived from their seismic and non-seismic
constraints can be obtained by modification of their masses. To do this, we write down six equations (for 
luminosity, radius and $D_{02}=\delta \nu_{02}/6$ of the each component) similar to equation (4) and solve them to 
find $M_{\rm A}$, $M_{\rm B}$, $\alpha_{\rm A}$, $\alpha_{\rm B}$, $X$ and $t$ for  fixed $Z$. 
These equations are not independent and therefore we get a set of solutions.
Some of these solutions are given in Table 3, for $Z=0.025$ and $Z=0.020$. For comparison, the values found from the 
observed data are also given (for  $\alpha$ Cen A, Bazot et al. (2007) gives $D_{0}=1.15\pm0.07 \mu$Hz). Different combinations of $M_{\rm A}$ and $M_{\rm B}$ are possible.
In Fig. 9, $M_{\rm B}$ is plotted against $M_{\rm A}$. The closest masses to the observed masses are
$M_{\rm B}$=0.922 $ \rm M_{\odot}$ and $M_{\rm A}$=1.115 $ \rm M_{\odot}$, for $Z=0.025$. Whereas $M_{\rm B}$ is less than the observed value by 0.012 $ \rm M_{\odot}$, $M_{\rm A}$ is greater 
than the observed value by 0.010 $ \rm M_{\odot}$. 
We notice two common properties of these models: \\
(i)the age of the system is about 5.2-5.3 Gyr, \\
(ii) for each solution, except the one given in the first and the second rows,
the convective parameter (fifth column in Table 3) of $\alpha$ Cen A is smaller  than that of 
$\alpha$ Cen B. This result is in good agreement with the result given in fig. 5 of Y{\i}ld{\i}z (2007).

The large separations of the models of $\alpha$ Cen A and B whose radii are fitted to the observed radius
are greater than the observed large separations. In order to fit the large separations of the models to the
values inferred from asteroseismology, the radii of $\alpha$ Cen A and B should be 
$R_{\rm A}$=1.230 $ \rm R_{\odot}$ and $R_{\rm B}$=0.875 $ \rm R_{\odot}$.

The adiabatic oscillation frequencies of the best fitted models for $\alpha$ Cen A and B with masses $1.115$ $\rm M_{\odot}$ and $0.922$ $\rm M_{\odot}$, respectively,
are given in Table 4. The fundamental properties of these models are presented in the sixth and the seventh rows of Table 3.


\section{Conclusions}
{
Using solar models and models of $\alpha$ Cen  A and B, the seismic properties of these 
stars are investigated in detail. Because of inconsistent results on the age of the $\alpha$ Cen system from the 
classical and seismic constraints, we test how well the small separation indicates age.
In this context, we consider two expressions in place of the customarily used form of 
the small separation 
between the frequencies ($\delta \nu_{02}$ or $D_0$) : \\
(i) the second difference, $\delta \nu_{012} = (\nu_{\rm n,0} - 2 \nu_{\rm n,1} + \nu_{\rm n,2})$; and\\
(ii) the frequency ratio ($\nu_{\rm n,0}/\nu_{\rm n,2}$).   \\
For the Sun, these three expressions give consistent results: the solar models with age of 4.9-5.0 Gyr 
and with $Z_0\approx 0.0165$ are in better agreement with the helioseismic data than the solar model with $Z_0 = 0.0165$ and age 4.6 Gyr.
For $\alpha$ Cen A and B the situation is complicated: although $\delta \nu_{02}$ gives about 5.7-5.9 Gyr for the 
age of the system (nearly the TAMS of $\alpha$ Cen A), we deduce from the comparison of the observed and model 
values of $\delta \nu_{012}$ that $\alpha$ Cen A and B are not so much evolved.

The expression we use for the small separation, $\delta \nu_{012}$,
has several important advantages over the customarily used expression $\delta \nu_{02}$:\\
(i) the resolution of the seismic HR diagram with $\delta \nu_{012}$ for the massive solar-like stars (e.g., $\alpha$ Cen A) is higher
than that with $\delta \nu_{02}$;\\
(ii) the evolutionary track of a star in the new seismic HR diagram is in agreement with its evolutionary track 
in the classical HR diagram (luminosity-effective temperature); and\\
(iii) $\delta \nu_{012}$ is nearly independent of $n$.\\

$\delta \nu_{012}$ deserves more detailed investigation in order to judge whether it has more specific
advantages (Christensen-Dalsgaard, private communication).

From the three equations for $D_{ll'}=\delta \nu_{ll'}/(4l+6)$ for $\alpha$ Cen A and B, we obtain solutions for 
initial heavy element abundance, age and mass of each star: for $\alpha$ Cen B, 
$Z_0=0.020$, $t=3.50$ Gyr and $M_{\rm B}=1.006$M$_\odot$; for $\alpha$ Cen A,  $t=3.50$ Gyr and $M_{\rm A}=1.19$M$_\odot$.
$Z_0$ value found for $\alpha$ Cen A is not a reasonable value. Therefore, we adopt $Z_0=0.020$.
The ratio of masses ($M_{\rm A}/M_{\rm B}$) is the same as the mass ratio derived from the observation (Pourbaix et al. 2002).
The heavy element abundance at the surface of model of $\alpha$ Cen B with diffusion (SIS2) is $Z_{\rm s}=0.0179$. 
This means that the average overabundance relative to solar ($Z_{\rm s}=0.0122$, AGS2005) is 0.17 dex.

Using the non-seismic constraints and $\delta \nu_{02}$ of $\alpha$ Cen A and B we find 
 from the solution of six equations that the age of the system is about 5.2-5.3 Gyr.
We also confirm that the mixing length parameter of $\alpha$ Cen A is smaller than that of $\alpha$ Cen B.
This is consistent with the result given in fig. 5 of Y{\i}ld{\i}z (2007), which shows that
the mixing length parameter of $\alpha$ Cen A is a decreasing function of time. 
Another interesting result we find is that the ages of the Sun and the $\alpha$ Cen system is very close to each other, about 5 Gyr.

The models of $\alpha$ Cen A and B, that have the same luminosity and radius as the observed values,
have large separations greater than the observed values.
If $R_{\rm A}$=1.230 $ \rm R_{\odot}$ and $R_{\rm B}$=0.875 $ \rm R_{\odot}$, 
the large separations of the models are in good agreement with the                                                 
values inferred from asteroseismology. 
  
In comparing the old (Grec et al. 1983) and the recent (Chaplin et al. 1999) seismic solar data and find 
that the level of accuracy of the seismic data of  $\alpha$ Cen  A and B is very similar to that of the old solar data. 
We hope that the situation will be much better in the near future than the present.

}
\section*{Acknowledgments}

J. Christensen-Dalsgaard is acknowledged for helpful discussions and for providing his adiabatic pulsation code.
This work is supported by The Scientific and 
Technological Research Council of Turkey (T\"UB\.ITAK).

\def \apj#1#2{ApJ,~{#1}, #2}
\def \aj#1#2{AJ,~{#1}, #2}
\def \astroa#1{astro-ph/~{#1}}
\def \pr#1#2{Phys.~Rev.,~{#1}, #2}
\def \prt#1#2{Phys.~Rep.,~{#1}, #2}
\def \rmp#1#2{Rev. Mod. Phys.,~{#1}, #2}
\def \pt#1#2{Phys.~Today.,~{#1}, #2}
\def \pra#1#2{Phys.~Rev.,~{ A}~{#1}, #2}
\def \asap#1#2{A\&A,~{#1}, #2}
\def \aandar#1#2{A\&AR,~{#1}, #2}
\def \apss#1#2{~Ap\&SS,~{{#1}}, #2}
\def \asaps#1#2{A\&AS,~{#1}, #2}
\def \arasap#1#2{ARA\&A,~{#1}, #2}
\def \pf#1#2{Phys.~~Fluids,~{#1}, #2}
\def \apjs#1#2{ApJS,~{#1}, #2}
\def \pasj#1#2{PASJ,~{#1}, #2}
\def \mnras#1#2{MNRAS,~{#1}, #2}
\def \ibvs#1{IBVS,~No.~{#1}}

\label{lastpage}

\end{document}